\documentclass{IEEEtran4PSCC}
\ifCLASSINFOpdf
   \usepackage[pdftex]{graphicx}
\else
   \usepackage[dvips]{graphicx}
\fi
\usepackage{url} 

\usepackage{standalone}
\usepackage[cmex10]{amsmath}
\usepackage{amssymb}
\usepackage{booktabs}
\usepackage{arydshln}
\usepackage{pgfplots}
\usetikzlibrary{intersections}
\usepgfplotslibrary{fillbetween}

\usepackage{myPlotStyle}
\makeatletter
\let\old@ps@headings\ps@headings
\let\old@ps@IEEEtitlepagestyle\ps@IEEEtitlepagestyle
\def\psccfooter#1{%
    \def\ps@headings{%
        \old@ps@headings%
        \def\@oddfoot{\strut\hfill#1\hfill\strut}%
        \def\@evenfoot{\strut\hfill#1\hfill\strut}%
    }%
    \def\ps@IEEEtitlepagestyle{%
        \old@ps@IEEEtitlepagestyle%
        \def\@oddfoot{\strut\hfill#1\hfill\strut}%
        \def\@evenfoot{\strut\hfill#1\hfill\strut}%
    }%
    \ps@headings%
}
\makeatother

\renewcommand{\baselinestretch}{0.98}
\begin{document}
\title{Neural Operators for Power Systems: \\A Physics-Informed Framework for Modeling Power System Components}

%% To specify the authors when (number of affiliations <= 2)
\author{
\IEEEauthorblockN{Ioannis Karampinis*, Petros Ellinas, Johanna Vorwerk and Spyros Chatzivasileiadis }
\IEEEauthorblockA{Technical University of Denmark, Wind and Energy Systems, Kongens Lyngby, Denmark\\
iokar@dtu.dk
% Name of the organization, acronyms acceptable\\
% City, Country\\
% \{email author n.1, email author n.2\}@domain (if desired)
}
}

%% To specify the authors when (number of affiliations > 2)
% \author{\IEEEauthorblockN{Author n.1\IEEEauthorrefmark{1},
% Author n.2\IEEEauthorrefmark{2},
% Author n.3\IEEEauthorrefmark{3}, 
% Author n.4\IEEEauthorrefmark{3} and
% Author n.5\IEEEauthorrefmark{4}}
% \IEEEauthorblockA{\IEEEauthorrefmark{1} Department Name of Organization A\\
% Name of the organization A,
% Address A\\ Emails if wanted}
% \IEEEauthorblockA{\IEEEauthorrefmark{2} Department Name of Organization B\\
% Name of the organization B,
% Address B\\ Emails if wanted}
% \IEEEauthorblockA{\IEEEauthorrefmark{3} Department Name of Organization C\\
% Name of the organization C,
% Address C\\ Emails if wanted}
% \IEEEauthorblockA{\IEEEauthorrefmark{4}Department Name of Organization D\\
% Name of the organization D,
% Address D\\ Emails if wanted}
% }

\maketitle

\begin{abstract}
Modern power systems require fast and accurate dynamic simulations for stability assessment, digital twins, and real-time control, but classical ODE solvers are often too slow for large-scale or online applications. We propose a neural-operator framework for surrogate modeling of power system components, using Deep Operator Networks (DeepONets) to learn mappings from system states and time-varying inputs to full trajectories without step-by-step integration. To enhance generalization and data efficiency, we introduce Physics-Informed DeepONets (PI-DeepONets), which embed the residuals of governing equations into the training loss. Our results show that DeepONets, and especially PI-DeepONets, achieve accurate predictions under diverse scenarios, providing over 30 times speedup compared to high-order ODE solvers. Benchmarking against Physics-Informed Neural Networks (PINNs) highlights superior stability and scalability. 
%An open-source toolbox is released to support adoption and reproducibility. 
Our results demonstrate neural operators as a promising path toward real-time, physics-aware simulation of power system dynamics.
\end{abstract}


% Use this to place sponsorships
\thanksto{\noindent  This work was supported by the ERC Starting Grant VeriPhIED, Grant Agreement 949899, and the ERC Proof of Concept PINNSim, Grant Agreement 101248667, both funded by the European Research Council. Submitted to the 24th Power Systems Computation Conference (PSCC 2026).}

\section{Introduction}
Modern power systems are increasingly stressed by high penetrations of converter‐interfaced resources, fluctuating loads, and more aggressive market dispatch, all of which demand faster and yet accurate dynamic simulations \cite{Zhang2025GridCloud}. Traditional time‐domain simulations for ordinary differential equations (ODEs) that describe power systems, e.g.\ swing equations for synchronous machines or dynamics of power electronic converters, provide reliable accuracy but become prohibitively slow \cite{Milano2010PowerScripting} when applied to large networks or real-time tasks including online stability assessment, digital-twin operation, and closed-loop control design.

To address these challenges, the power systems community has turned to machine learning (ML) surrogates. Recurrent networks such as Long Short-Term Memory(LSTM) Networks \cite{Li2020Machine-learning-basedDynamics} and neural ODEs \cite{Xiao2022FeasibilityModeling} have been used to regress trajectories directly, but their reliance on large datasets generated by costly numerical solvers limits practicality. Physics-Informed Neural Networks (PINNs) integrate system equations into the training loss, reducing data requirements \cite{Karampinis2025ToolboxComponents}. However, vanilla PINNs, composed of a single Neural Network, are generally formulated for specific initial conditions and parametric evolution equations \cite{Stiasny2024PINNSim:Networks}. Moreover, they may accumulate inaccuracies over long integration horizons \cite{Wang2023Long-timeDeepONets}, which could limit their suitability for real-time operation.

Neural operator architectures offer a promising alternative, bypassing the need for sequential time-stepping in solving the governing ODEs. The Deep Operator Network (DeepONet) \cite{Lu2021LearningOperators} learns mappings between function spaces, acting as a continuous-time surrogate that can evaluate the response of a dynamical system at arbitrary time points without re-integration. Once trained, inference is nearly instantaneous. The learned operator can act as a rapid simulator—predicting future states from candidate control actions for real-time control loops, or as a digital twin for simulation, diagnostics, and analysis. Even though recent studies have demonstrated operator learning for power system components \cite{Arowolo2025ExploringSimulation} and for Multi-Input-Multi-Output control frameworks \cite{deJong2025DeepControl}, these efforts are mainly data-driven, with limited robustness and generalization capabilities..
 
To address this gap, this work explores Physics-Informed DeepONets (PI-DeepONets)\cite{Goswami2023Physics-informedNetworks} for non-autonomous power system components with time-varying inputs. By embedding governing physics into operator learning, we aim to combine data-driven speed with improved generalization, offering accurate and efficient surrogates for real-time simulation, analysis, and control.

To the best of our knowledge, this is the first application of Physics-Informed Neural Operators to power system components subject to time-varying inputs. The main contributions of the presented work are:
\begin{itemize}
\item \textbf{Neural‐Operator Framework for Power System Dynamics.} We propose one of the first applications of data-driven operator learning surrogate modeling framework to directly map initial conditions and time-varying inputs (e.g., mechanical torque, voltage set-points, fault events) directly to system states. The framework encodes time-varying inputs through sensor sampling, enabling trajectory evaluation at any query time without the need for step-by-step numerical integration.

\item \textbf{Physics‐Informed Operator Priors.} By embedding the governing differential‐equation residuals into the loss function, our PI-DeepONets \cite{Goswami2023Physics-informedNetworks} enforce physical constraints (e.g. swing‐equation and voltage‐dynamics), improving generalization to unseen operating points and reducing required training data. To the best of our knowledge, this is the first application of Physics-Informed Neural Operators in power systems.
%\item \textbf{Comprehensive Benchmark with Autoregressive Physics‐Informed Neural Networks(PINNs).} We perform head‐to‐head comparisons against autoregressive PINNs—where a PINN is trained to predict one time‐step ahead and then rolled out—to quantify accuracy, stability, computational cost and generalization capability.
\item \textbf{Speedup and Generalization Study.} We demonstrate that our operator models achieve at least a 30× speedup in inference time compared to high‐order ODE solvers across diverse component types, all while maintaining high accuracy.
\end{itemize}

% \textbf{Code Release:} 
Furthermore, we provide the implementation of our framework, including dataset generation, training routines, and evaluation scripts, to support adoption and ensure reproducibility of our results. The code is publicly available on GitHub: \url{https://github.com/radiakos/PowerDeepONet}

\section{Deep Neural Operators for Time-Varying Input Dynamical Systems}
\subsection{Problem Formulation}
Power system components exhibit dynamic behavior that can be mathematically represented using ODEs. These equations describe the evolution of state variables over time and are typically framed as initial value problems (IVPs). An IVP refers to finding a function $x(t)$ solution to a differential equation subject to given initial conditions and its external inputs. The uniqueness of the solution is ensured by the Picard-Lindelof theorem, assuming that the function $f$ governing the ODE is Lipschitz continuous. The IVP and its solution $x(t)$ are formulated as:
\vspace{-0.3em}
\begin{subequations}
\begin{align}\label{odes}
\frac{d}{dt} \mathbf{x(t)} = f (t, \mathbf{x(t)}, \mathbf{u(t)}, \boldsymbol{\mu}), \quad \mathbf{x}(t_0) = \mathbf{x}_0,
\\
\mathbf{x(t)} = \mathbf{x}_0 + \int_0^t f\big(\tau,\mathbf{x}(\tau), \mathbf{u}(\tau), \boldsymbol{\mu} \big)\, d\tau,
\label{odes2}
\end{align}
\end{subequations}

\vspace{-0.3em}
where $\mathbf{x}(t)$ represents the state vector that evolves over time $t$, $f$ is a given function that maps the system parameters to the states, vector $\mathbf{u(t)}$ represents the time-varying inputs of the system, vector $\boldsymbol{\mu}$ includes the system parameters that are to be determined, and $ \mathbf{x}_0 $ represents the initial conditions. For readability, the time argument $(t)$ is dropped from $\mathbf{u(t)}$, with the understanding that $\mathbf{u}$ is time-dependent.

The objective is to determine the function $\mathbf{x}(t;\mathbf{x_0},\mathbf{u},\boldsymbol{\mu})$ that fulfills \eqref{odes} and describes the system’s time-domain response. This function represents the unique evolution of the system over a time horizon $\mathbf{T}$ for the given initial conditions and time-varying input, referred to as a trajectory. Yet, because the problem is nonlinear, obtaining a closed-form expression for $\mathbf{x}(t)$ is not feasible, so we must rely on approximations: 

\vspace{-0.5em} 
\begin{equation}
\hat{\mathbf{x}}(t;\mathbf{x_0},\mathbf{u},\boldsymbol{\mu}) \approx  \mathbf{x}(t;x_0,\mathbf{u},\boldsymbol{\mu}).
\end{equation}

%where $u \in V$ (a discretized subset of $C([0,T])$) is the input signal, and $ \mathbf{x}(t) : [0,T] \to \mathbf{R}^K$ is the solution of system serving as the output signal. \\

%However, in practice the input function is unknown, and we rely on approximations $\hat{\mathbf{u}}$ . Our goal is to approximate the future state of a non-autonomous dynamical system, requires the pre knowledge of the external input. The proposed multi-step prediction solution can be combined with model predictive control actions, either with existing established solutions or a machine-learning based predictor of the external inputs' evolution. During this work we will consider $\hat{\mathbf{u}} \equiv {\mathbf{u}}$ and stress the ability of the operator to approximate the behavior of the system under various inputs.
\vspace{-0.3em}
In practice the input function \(\mathbf{u}\) is often unknown or partially known, and we rely on an approximation \(\hat{\mathbf{u}}\). Since our objective is to approximate the future state of a non-autonomous, i.e. interconnected,  dynamical system, knowledge of the external input is required. The proposed multi-step prediction framework can be combined with model predictive control (MPC) strategies, either leveraging existing established approaches or employing an ML-based predictor for the evolution of 
external inputs. In the presented work, we %make the simplifying assumption \(\hat{\mathbf{u}} \equiv \mathbf{u}\) and 
stress the ability of the operator to approximate the behavior of the system under various inputs, having access to the same knowledge as traditional ODE solvers.

%Additionally, the operator requires the discretizion of the function ${\mathbf{u}}$ in a number m+ 1 of sensors within [0,T]  in predefined locations. This allows to parse it as an input in the branch net. The number of sensors within the desired time horizon should be sufficient enough to have increase entropy that enables the branch net to capture its core characteristics. 

Since no analytical solution is available for the ODEs, the system response must be approximated through discretization, i.e., by evaluating the dynamics at specific time instances. %Similarly, the operator requires discretization of the input function \(\mathbf{u}\) at \(m+1\) sensor locations \(\{t_i\}_{i=0}^m \subset [0,T]\). The resulting vector $\big(u(t_0),\, u(t_1),\, \dots,\, u(t_m)\big)$ time traces are provided as input to the branch network. The number and placement of sensors within the time horizon must be chosen such that the discretized time traces preserve sufficient information about \(\mathbf{u}\). In particular, \(m\) should be large enough to ensure that the induced sampling entropy of \(\mathbf{u}\) is sufficient for the branch network to capture its core characteristics. 
\subsection{Classical Ordinary Differential Equations Solvers}

Traditional numerical solvers, such as Runge–Kutta schemes, are commonly used for solving ODEs. These methods approximate the solution trajectory over a specified time interval T by iteratively computing discrete updates based on derivative
evaluations for small time steps in specific time instances. Increasing the number of data points, i.e., using smaller step sizes, generally improves the accuracy of the approximation but also increases the number of iterations required and subsequently the computational cost. This iterative process can be computationally expensive, especially for complex systems with many state variables, stiff equations. 

These limitations motivate the exploration of operator-learning approaches, such as DeepONets, which directly approximate the solution operator.
\subsection{Deep Operator Network}
%DeepONet consists of two sub-networks, the Branch Net and the Trunk Net. Branch Net encodes the input functions at a ﬁxed number of sensors  $x_i \in [1,...,m]$. Trunk Net encodes the locations for the output functions. DeepONet aims to directly map the system states, time-varying input functions (mechanical torque, voltage set-points, fault events) and the location time to the resulting state at this time. In our approach DeepONet approximates the state of the non-autonomous dynamical system using a future approximation of the time-varying input, mapping the initial conditions of the system, the discretized time-varying input in a given number of sensors, and the independent variable t within the horizon T. 

DeepONet is composed of two sub-networks: the \emph{Branch Net} and the \emph{Trunk Net}, which are both introduced in the following.

\subsubsection{Branch Net} encodes the input function(s) at a fixed number of sensors $x_i \in [1,\ldots,m]$ and outputs a set of coefficients $b_k(\mathbf{x})$. In this context, sensors correspond to discrete sampling locations of the input signal
%The Branch Net can be consisted of a single Neural Network in the case of Unstacked DeepONet, but also multiple in that of Stacked DeepONet. For the later, each sensor data is feeded in its own Neural Network. Each of the two architecture has its advantages, based on the accuracy of the DeepONet and its generalization ability, as they encode in a different way the information from the input functions. In this work we examine the performance of the Unstacked DeepONet and an alternative of the Stacked DeepONet, which we call N-Stacked DeepONet. In the case of N-Stacked DeepONet we deploy as many networks as the two branch networks, one to encode the initial condition and one to encode the time varying input. This approach has a potential for modeling multi-stage or hierarchical operators, where the external input is encoded by a different encoder, or embeds an whole approximator for the time-varying input. 
Similarly to the traditional ODE solvers, the operator requires discretization of the time-varying input function \(\mathbf{u}\) at \(m+1\) sensor locations \(\{t_i\}_{i=0}^m \subset [0,T]\). The resulting vector $\big(u(t_0),\, u(t_1),\, \dots,\, u(t_m)\big)$ time traces are provided as input to the Branch Net. The number and placement of sensors-samples within the time horizon must be chosen such that the discretized time traces preserve sufficient information about \(\mathbf{u}\). In particular, \(m\) should be large enough to ensure that the induced sampling entropy of \(\mathbf{u}\) is sufficient for the Branch Net to capture its core characteristics. 

\begin{figure}[tbp]
  \centering
  \includegraphics[width=\linewidth]{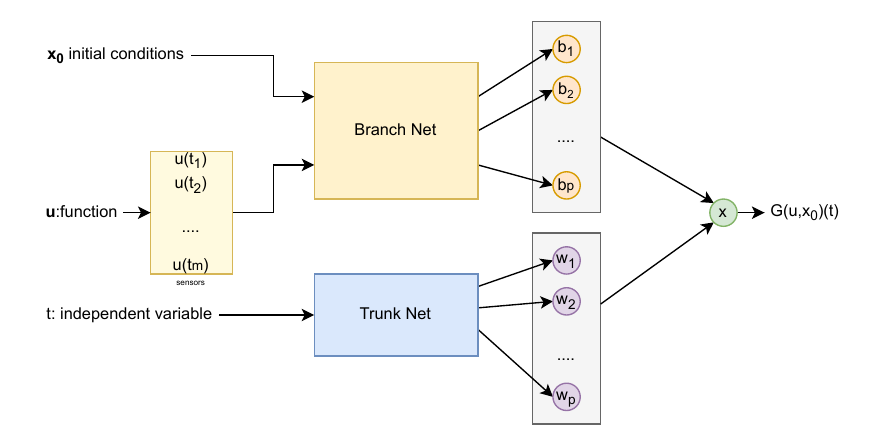}
 
  \caption{Traditional Unstacked DeepONet architecture}
  \label{fig:Unstacked}
  \vspace{-1em}
\end{figure}

\begin{figure}[tbp]
 \centering
 \includegraphics[width=\linewidth]{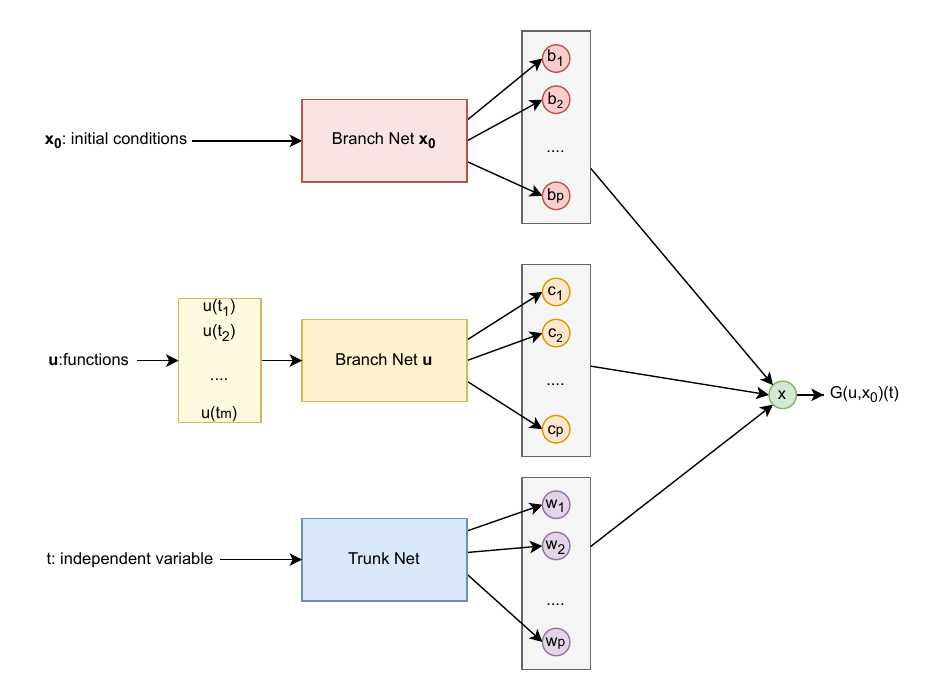}
 \caption{Stacked-N DeepONet architecture}
 \label{fig:Stacked-N}
 \vspace{-1em}
\end{figure}

The Branch Net can consist of either a single Neural Network(NN), called \emph{Unstacked DeepONet} as presented in Fig.~\ref{fig:Unstacked}, or multiple networks, called \emph{Stacked DeepONet}. In the latter case, each sensor value is processed by its own Neural Network. Both architectures have distinct advantages in terms of approximation accuracy and generalization, since they encode the information from the input functions in different ways. In this work, we evaluate the performance of the Unstacked DeepONet and an alternative variant of the Stacked DeepONet, which we refer to as the \emph{Stacked-N DeepONet} and shown in Fig.~\ref{fig:Stacked-N}. In the Stacked-N architecture, we employ two dedicated branch networks, each corresponding to a different type-nature of input: one to encode the initial conditions and another one to encode the time-varying input. This modular design is particularly promising for modeling multi-stage or hierarchical operators, where the external input \(\mathbf{u}\) can be handled by a specialized encoder or even by an entire surrogate approximator.
\subsubsection{Trunk Net} encodes the coordinates of the output function, typically the temporal or spatio-temporal evaluation points, and outputs a set of basis functions $\tau_k(\mathbf{t})$. 

The output of the DeepONet is obtained by combining the latent representations of the two sub-networks as follows: let $G$ denote the operator mapping the input functions $\mathbf{x}$ and the query point $\mathbf{t}$, so that the desired solution is $G(\mathbf{x})(\mathbf{t})$. In practice, DeepONet approximates this operator through an expansion of the form 
\vspace{-0.8em} 
\begin{equation} \label{eq:deeponet}
G(\mathbf{x})(\mathbf{t}) \approx \sum_{k=1}^p b_k(\mathbf{x}) \, \tau_k(\mathbf{t}),
\end{equation} 

\vspace{-0.3em} 
where $b_k(\mathbf{x})$ is the output of the Branch Net and $\tau_k(\mathbf{t})$ is the output of the Trunk Net.
Based on the universal approximation theorem for operators \cite{Lu2021LearningOperators}, DeepONet architectures with a sufficient number of sensors and basis functions can approximate nonlinear continuous operators to an arbitrary degree of accuracy.


%\subsubsection{Trunk Net}
%family of functions
%learns an operator that maps input functions output functions.

%Input branch: encodes the function (you can give it sampled values or parameters).

%Output branch: takes the query time t. Network outputs f(t).
%Very fast at inference once trained. 

In our setting, DeepONet $G$ aims to directly map the initial system state $\mathbf{x_0}$, the time-varying input function(s) $\mathbf{u}$(e.g., mechanical torque, voltage set-points, fault events), and the query time $\mathbf{t}$ to the resulting system state at $\mathbf{t}$. Thus, DeepONet provides a surrogate for the solution operator of a non-autonomous dynamical system, by mapping the initial conditions, the discretized time-varying input, and the independent variable 
$\mathbf{t}$. Importantly, it should satisfy the underlying dynamical relation without the need for explicit numerical re-integration of the governing equations. Consequently, it is expected to fulfill equation \eqref{odes2} as follows:
\vspace{-0.5em} 
\begin{equation} \label{eq:deeponettime} 
G(\mathbf{u}, \mathbf{x_o})(t) = \mathbf{x_o} + \int_0^t f\big(\tau,G(\mathbf{u}, \mathbf{x_o})(\tau) , \mathbf{u}(\tau), \boldsymbol{\mu} \big)\, d\tau. 
\end{equation} 
%$G(u, x_o)$ satisfies \[ G(u, x_o)(t) = x_0 + \int_0^t f\big(\tau,G(u, x_o)(\tau) , \mathbf{u}(\tau), \mu \big)\, d\tau. \]
\vspace{-1.5em} 
\subsection{Training the Solution Operator DeepONet}
DeepONets can be trained on labeled simulated data of input--output pairs, allowing them to generalize to unseen inputs when the training set sufficiently represents the target domain. 
Training involves iteratively updating the model parameters to minimize the discrepancy between predictions and desired outputs. In our setting, such data can be generated using ODE solvers. Trajectories are simulated and transformed into the required format. The resulting dataset, defined over the input domain $\Omega_d$ with $N_d$ samples has the following format: $\{(x_0^{(i)},\mathbf{u}^{(i)}, t^{(i)}), x^{(i)}\}_{i=1}^{N_d}$. In case of using the mean squared error (MSE) as the loss function between the ground truth state $x^{(i)}$ and the DeepONet output $\hat{x}^{(i)}$, the trainable parameters are updated according to:
\vspace{-0.8em} 
\begin{equation} \label{eq:data_loss}
\mathcal{L}_{\text{data}} = \frac{1}{N_d} \sum_{i=1}^{N_d} \lVert \hat{x}^{(i)} - x^{(i)} \rVert^2 .
\end{equation}
\vspace{-0.5em} 
\subsubsection*{Physics-Informed DeepONet}
The structure of the proposed DeepONet allows for Physics-Informed (PI) training when the underlying system of ODEs is available. As proposed in \cite{Raissi2019Physics-informedEquations} and implemented in \cite{Karampinis2025ToolboxComponents}, we exploit automatic differentiation (AD) to incorporate physical constraints directly into the loss function. In more detail, AD provides the derivative of the DeepONet's output with respect to the query point $\mathbf{t}$(given as input), corresponding to the left-hand side of \eqref{odes}, while the right-hand side can be evaluated either using the ground-truth $x^{(i)}$ from the simulated dataset or the DeepONet prediction itself. The resulting PI loss based on simulated data $\mathbf{x}^{(i)}$ is expressed as:
\vspace{-1.2em}
\begin{equation} \label{eq:physics_data_loss}
\mathcal{L}_{\text{phy data}} = \frac{1}{N_d} \sum_{i=1}^{N_d} \left| \!\!
\underbrace{\frac{d \hat{\mathbf{x}}^{(i)}}{d t}}_{\text{\tiny \begin{tabular}{c}Derivative of\\DeepONet output \end{tabular} }} \!\!-\!
\underbrace{f\left(t^{(i)}, \mathbf{x}^{(i)}, \mathbf{u}, \boldsymbol{\mu}\right)}_{\text{\tiny Source term evaluation}} 
\right|^2. 
\end{equation}

DeepONet serves as a solution operator and should satisfy the governing physics equations at any arbitrary point within the training domain. When the physics loss is formulated using only the DeepONet approximation, both sides of \eqref{odes} depend solely on the DeepONet’s predictions $\hat{\mathbf{x}}^{(i)}$. This allows us to introduce additional evaluation points, referred to as collocation points, where the physics residual is enforced using as input to $f$ the DeepONet outputs  $\hat{\mathbf{x}}^{(i)}$ and not $\mathbf{x}^{(i)}$ like in \eqref{eq:physics_data_loss}, as follows:\vspace{-0.8em}
\begin{equation} \label{eq:physics_col_loss}
\mathcal{L}_{\text{phy col}} \!= \!\frac{1}{N_d} \sum_{i=1}^{N_d} \left|\!\!
\underbrace{\frac{d \hat{\mathbf{x}}^{(i)}}{d t}}_{\text{\tiny \begin{tabular}{c}Derivative of \\ DeepONet output\end{tabular}}} \!\!\!- \!\!\!\!\!
\underbrace{f\!\left(t^{(i)},\hat{\mathbf{x}}^{(i)}, \mathbf{u}, \boldsymbol{\mu}\right)}_{\text{\tiny\begin{tabular}{c} DeepONet approximation term evaluation\end{tabular}}} \!
\right|^2.
\end{equation}
% \textcolor{red}{Notice that in \eqref{eq:physics_col_loss}, the input to $f$ is $\hat{\mathbf{x}}^{(i)}$ and not $\mathbf{x}^{(i)}$ like in \eqref{eq:physics_data_loss}. This means that in \eqref{eq:physics_data_loss} $f$ offers the actual ground truth derivative, while in \eqref{eq:physics_col_loss} the derivative is estimated based on the output of the NN.}

\section{A Training Pipeline for DeepONets}
The main goal of this work is to create a reliable and efficient pipeline for training DeepONets to model non-autonomous dynamic components and demonstrate their performance.. The pipeline is built to be fully parameterizable and highly adaptable at each stage. The following sections present the core steps of the methodology and explain the reasoning behind the choices made at each step.

\subsection{Setting up the ODEs with the Time-Varying Input}
Initially, we define the set of ODEs in \eqref{odes} that govern the desired non-autonomous power system dynamic component without replacing any value. The system parameters $\mu$ in \eqref{odes} are considered static and are stored separately. This approach allows testing components over a range of parameters, such as varying the inertia of a synchronous machine. The resulting ODEs are then employed for the dataset generation by the ODE solver and for the PI training. 

\subsection{Dataset Generation}
The second phase includes the generation of the datasets where one dataset contains the simulated trajectories and the other the collocation points. During the training, we expose the ML model to a limited training region, while for the testing we can assess its ability to capture the behavior of the examined system both within the same input domain, and also stress its generalization ability. This results into a limited domain, i.e. limited samples within a range of values, for the initial conditions, a specific time horizon and a family of external input signals with specified characteristics:
\begin{itemize}
    \item \textbf{Sampling the initial conditions:} We seek to approximate the system’s behavior over a defined input domain $ \Omega_d$. To do so, we first specify the boundaries of this domain and determine how many distinct samples to take for each state. Different sampling methods can be employed, such as uniform, random, Latin Hypercube sampling (LHS) \cite{McKay2000ACode} and others. LHS is preferred as a stratified sampling technique to ensure that all portions of the input space are sampled, allowing for comprehensive coverage of the input space with fewer samples.
    \item \textbf{Time horizon:} The surrogate model must be trained over a finite time window $[0, T]$, where $T$ is chosen according to the needs of the study. Additionally, the time intervals within the time window are to be defined based on the user's needs.
    \item \textbf{External input signals:} In addition to the initial conditions and the time horizon, the dataset must incorporate a representative family of time-varying input, swhose characteristics are assumed to be known or predefined. These signals can be polynomial, sinusoidal, stochastic, or designed to mimic realistic operating conditions. Their characteristics (e.g., amplitude, frequency, bandwidth) are selected within predefined bounds to ensure that the model learns to generalize across a sufficiently diverse set of operating scenarios.
\end{itemize}

Having defined all the above, the ODE solver produces a number of simulated trajectories. These are then further processed to obtain the needed input-output format. The resulting labeled data is exploited during training through the labeled losses \eqref{eq:data_loss} and \eqref{eq:physics_data_loss}. In contrast, for the collocation points dataset, only the construction of the input format is necessary, and therefore the use of the ODE solver is not required.
\subsection{Deep Operator Neural Network (DeepONet) Architecture}

Depending on the desired approach, we must define the architecture of DeepONet to be implemented. The complexity of the examined power system component directly affects the number of hidden layers and the neurons per layer in the Branch and Trunk Nets. The more complex the examined system is, the more neurons and/or layers must be included. The width of the input layer of the Branch Net is determined by the number of the initial conditions and the sensors used, while for the Trunk Net has a single scalar input corresponding to the evaluation time t. The width of the output layer of both the Branch Net and the Trunk Net is a shared latent dimension $p$. 
The Branch Net maps the initial conditions and external inputs to $B(u)\in\mathbb{R}^p$, while the Trunk Net maps the evaluation time $t$ to $T(t)\in\mathbb{R}^p$. 
Their inner product forms the system response. 
The choice of $p$ depends on the complexity of the examined component: higher nonlinearity or richer dynamics require larger $p$, while simpler systems can be modeled with smaller values. This selection is typically treated as part of the hyperparameter tuning process.

\subsection{Training of DeepONet}
Having defined both the training datasets and the DeepONet's architecture, the next phase includes the training of the DeepONet, which is the optimization of the trainable parameters (i.e. the weights of neurons). The goal is to minimize the losses described in \eqref{eq:data_loss}, \eqref{eq:physics_data_loss} and \eqref{eq:physics_col_loss}. Since the loss term consists of different terms, a balance between the data points discrepancy and the physics loss must be found. Different strategies can be employed to optimize the total performance of the trained ML model, to provide reliable results with the minimal error. The total loss is a weighted sum of all the losses available, where the weights can be static or dynamically adaptable. The total loss will then be used by the selected optimizer to adjust the trainable parameters of the ML model over a repetitive process. 

\subsection{Testing: Performance Assessment of DeepONet}
The developed surrogate models are evaluated along two key dimensions: accuracy and computational efficiency.
\subsubsection{Accuracy}

Depending on the intended application, the required performance of the DeepONet may vary, e.g. higher accuracy may be needed near the steady state, during the initial transients, or uniformly across the entire time horizon. Accordingly, different sampling strategies for dataset generation and weighting schemes in the loss function can be employed to emphasize specific regions of interest.

In this work, we aim to train and evaluate a DeepONet surrogate model that performs uniformly well within the interval [0,T]. For testing, a number of labeled trajectories from the same input domain can be used. The model’s performance is evaluated using the Mean Absolute Error (MAE), the Mean Squared Error (MSE), and the Maximum Absolute Error (MaxAE) that captures the upper error bound. These metrics are computed both globally over the full time horizon and locally at each time step for all state variables.

While MAE provides a linear measure of the average magnitude of prediction errors, offering a direct interpretation of model's deviation, the MaxAE quantifies the largest observed absolute error, defining the model’s upper error bound.

\subsubsection{Time}

The computational efficiency of the developed DeepONet is assessed by comparing its inference time with the traditional numerical ODE solver. Unlike solvers that iteratively compute all intermediate steps, the DeepONet directly approximates system states at arbitrary time points, bypassing the integration process. This allows rapid evaluation of single points or entire trajectories in one forward pass. Moreover, the parallelism nature of Neural Networks allows efficient GPU utilization and larger batch sizes, allowing simultaneous evaluation of multiple trajectories. The resulting speedup of DeepONet inference time is measured against solver runtime, in both single- and multi-trajectory simulations, and also in single point evaluation.

\section{Case study}
This section demonstrates DeepONets ability to approximate the behavior of a power system's component with time-varying external input. Specifically, we present results for the two different aforementioned architectures of DeepONet, Unstacked and Stacked-N, as a surrogate for a 4th-order synchronous machine~(SM) connected to a bus with varying voltage. 
%This study considers the 4th-order representation found in \cite{SauerPOWERSTABILITY} as implemented in \cite{KarampinisToolboxComponents} and the parameters used are shown in Table~\ref{smparams} 
This study considers the following representation found in \cite{Sauer2017PowerToolbox}  and the parameters used are shown in Table~\ref{smparams} : 
\vspace{-0.8em}
\begin{multline}
\hspace{-1.4em}
% \noindent
%\scriptsize
\begin{bmatrix}
%\scriptsize
1 \\
\frac{2H}{\Omega_{B}} \\
T'_{do} \\
T'_{qo} 
\end{bmatrix}
\hspace{-0.3em}
\frac{d}{dt}
\hspace{-0.3em}
\begin{bmatrix}
\delta \\
\omega \\
E'_q \\
E'_d 
\end{bmatrix}
\hspace{-0.4em}= \hspace{-0.4em}
\begin{bmatrix}
%\scriptsize
\omega \\
P_m\!\!-\!\!E'_d I_d\! + \!E'_q I_q\! +\! (X'_q\!-\!X'_d)I_d I_q \!\!-\! \!D\omega \\
-E'_q - (X_d - X'_d)I_d + E_{fd} \\
-E'_d + (X_q - X'_q)I_q 
\end{bmatrix}\!\!\!
\end{multline}
% \end{align}
and 
\begin{align}
\begin{bmatrix}
(R_s \!+ \!R_e) & \!-(X_q\! + \!X_e) \\
(X'_d \!+ \!X_e) & (R_s \!+\! R_e)
\end{bmatrix}\!\!
\begin{bmatrix}
I_d \\ I_q \\ 
\end{bmatrix}
\!\!&=\!\!
\begin{bmatrix}
E_d' \!- \!V_s \!\sin(\delta \!- \!\theta_{vs}) \\
E_q'\! -\! V_s\! \cos(\delta \!- \!\theta_{vs}) \\
\end{bmatrix}.
\end{align}

%\begin{align}
%\begin{bmatrix}
%P_e \\ V_d \\ V_q \\ V_t \\ S_E(E_{fd})
%\end{bmatrix}
%&=
%\begin{bmatrix}
%E'_d I_d + E'_q I_q + (X'_q - X'_d)I_d I_q \\
%R_e I_d - X_{ep} I_q + V_s \sin(\delta - %\theta_{vs}) \\ 
%R_e I_q - X_{ep} I_d + V_s \cos(\delta - %\theta_{vs}) \\ 
%\sqrt{V_d^2 + V_q^2} \\
%0.098 * e^{(0.55*E_{fd})})
%\end{bmatrix}
%\end{align}

\begin{table}[t]
\centering
\caption{Synchronous Machine Parameters}
\begin{tabular}{ll}
\toprule
\textbf{Parameter} & \textbf{Value} \\
\midrule 
Damping factor, $D$ & 2 \\
Inertia constant, $H$ [s] & 5.06 \\
Stator resistance, $R_s$ [p.u.] & 0 \\
Direct-axis transient time constant, $T_d'$ [s] & 4.75 \\
Quadrature-axis transient time constant, $T_q'$ [s] & 1.6 \\
Direct-axis reactances, $X_d$, $X_d'$ [p.u.] & 1.25, 0.232 \\
Quadrature-axis reactances, $X_q$, $X_q'$ [p.u.] & 1.22, 0.715 \\
Line reactance, $X_{ep}$ [p.u.] & 0.1 \\
Line resistance, $R_e$ [p.u.] & 0 \\
System angular frequency, $\Omega_b$ [rad/s] & 314.159 \\ \bottomrule
\label{smparams}
\vspace{-2.5em}
\end{tabular}
\end{table}

\subsection{ Datasets of the Input Domain}
\subsubsection{Domain of Initial Conditions} 
Different initial conditions for the simulations are sampled from the domain of initial conditions $\Omega_{ic}$. The selected ranges are as follow: $\theta \in [-2, 2]$ rad, $\omega \in [-1, 1]$ rad/s, $E_d'$ = 0 pu, $E_q' \in [0.9, 1.1]$ pu, $R_F $ = 1 pu, $V_r $ = 1.105 pu, $E_{fd}$ = 1.08 pu, $P_{sv}$ = 0.7048 pu, and $P_m$= 0.7048 pu. In total, 2\,200 different sets of initial conditions were sampled within the state space using LHS, consisting of 1\,000 for the labeled training data, 1\,000 for the collocation points, and 100 for each of the labeled validation and test sets.

\subsubsection{Time-Varying External Input}
The external voltage inputs, defined by the magnitude $V_s$ and the phase angle $\theta_{vs}$, are modeled using either first or second-order ODEs. An alternative modeling approach was proposed in \cite{Wang2018ASimulation} and
employed in \cite{Stiasny2024PINNSim:Networks}, where the evolution of the external input is
represented in polar form. The two general approaches used in this work are formulated as follows:
\begin{equation}
\frac{dy}{dt} = -k(y - y_{\text{ref}}) + A_{\text{noise}}\sin(\omega_{\text{noise}} t),
\label{ode1}
\end{equation}
or
\begin{equation}\quad 
\frac{d^2 y}{dt^2} + 2\zeta\omega_n\frac{dy}{dt} + \omega_n^2(y - y_{\text{ref}}) = 0,
\label{ode2}
\end{equation}
where $k$ is the system gain, $y_{\text{ref}}$ the steady-state reference, $A_{\text{noise}}$ and $\omega_{\text{noise}}$ denote the amplitude and frequency of perturbations, and $\zeta$ and $\omega_n$ are the damping ratio and natural frequency, respectively.

Two configurations of external voltage dynamics are considered for the presented case study:
\begin{enumerate}
    \item \textbf{Slow response:} Both $V_s$ and $\theta_{vs}$ follow first-order dynamics defined in \eqref{ode1} with low-frequency perturbations.
    \begin{itemize}
        \item \textit{Voltage magnitude:} $k \in [1.0, 2.0]$, $A_{\text{noise}} \in [0.0, 0.03]$, $\omega_{\text{noise}} \in [0.5, 1.5]$, $V_s(0) \in [0.7, 1.2]$
        \item \textit{Phase angle:} $k \in [1.0, 3.0]$, $A_{\text{noise}} \in [0.0, 0.05]$, $\omega_{\text{noise}} \in [0.1, 0.5]$, $\theta_{vs}(0) \in [-0.5, 0.5]$
    \end{itemize}

    \item \textbf{Fast response:} Both quantities are modeled as second-order systems as in \eqref{ode2} with higher frequencies. 
    \begin{itemize}
        \item \textit{Voltage magnitude:} $\zeta \in [0.6, 0.8]$, $\omega_n \in [8.0, 12.0]$, $V_s(0) \in [0.7, 1.2]$, $\dot{V}_s(0) \in [-0.2, 0.2]$
        \item \textit{Phase angle:} $\zeta \in [0.2, 0.4]$, $\omega_n \in [10.0, 18.0]$, $\theta_{vs}(0) \in [-0.5, 0.5]$, $\dot{\theta}_{vs}(0) \in [-0.8, 0.8]$
    \end{itemize}
\end{enumerate}

\subsubsection{Generating Labeled and Collocation Datasets}

The numerical solver \texttt{RK45} from the \texttt{scipy.integrate} library is used to generate the simulated trajectories representing the transient behavior of the SM. The solver inputs the time-varying external signals, the initial conditions, and the time horizon. The latter is set to 1 s with an evaluation time step of 1 ms. The external input functions were sampled at the same evaluation points, resulting in 1\,000 sensors in total. The outputs correspond to the simulated trajectories of each SM state variable, which were subsequently processed to form the input–output pairs required for DeepONet training.

In contrast, the collocation dataset consists only of inputs. Specifically, it includes the evaluation points for the trunk network, the initial conditions, and the sampled external input vectors.

The final datasets have the following format: 
\begin{subequations}\label{eq:T1T2}
\begin{align}
\mathcal{X}_{data} & = \{\big(\mathbf{x}_0^{(i)},\mathbf{u}^{(i)}, t^{(i)}\big), \mathbf{x}^{(i)} \}_{i=1}^{N_{l}} \\
\mathcal{X}_{collocation} & =   \{\big(\mathbf{x}_0^{(i)},\mathbf{u}^{(i)}, t^{(i)}\big) \}_{i=1}^{N_{c}}  
\end{align}
\end{subequations}

The training dataset comprises 1\,000 labeled trajectories, each containing 1\,000 data points (in total, $1{\,}000{\,}000$ data points), along with $N_c = 1{\,}000{\,}000$ collocation points. The validation and testing datasets each contain 100 trajectories of identical dimensionality in time.

\subsection{DeepONet Architecture and Hyperparameter Training}

For both implementations, Unstacked and Stacked-N, the Branch and Trunk networks are fully connected neural networks consisting of three hidden layers with 64 neurons each and a hyperbolic tangent (\texttt{tanh}) activation function. The output layer of each network produces 64 latent features, whose inner product approximates the response of the dynamical component. The architecture was determined empirically by testing different hyperparameters, with the final configuration providing a favorable trade-off between accuracy and computational cost \& efficiency.

Training is performed using the SOAP optimizer \cite{Vyas2024Soap:Adam} with a learning rate of $3 \times 10^{-3}$ and a weight decay of 0.01. The models are trained for 10\,000 epochs using the Mean Squared Error (MSE) loss function. Early stopping is employed to prevent overfitting, with a patience of 200 epochs and a minimum improvement threshold of $1 \times 10^{-6}$. The static loss weights are set to $\lambda_d = 1$, $\lambda_{pd} = 1 \times 10^{-3}$, and $\lambda_{pc} = 1 \times 10^{-4}$.

All implementations were developed in PyTorch \cite{PyTorchFoundationPyTorch}. Weights \& Biases (WandB) \cite{Biewald2020ExperimentBiases} was used for experiment tracking and hyperparameter optimization, providing detailed insights into model training and testing performance. All experiments were conducted on the High-Performance Computing (HPC) cluster at the Technical University of Denmark, utilizing a 16-core Intel Xeon 6226R processor with 256 GB of RAM and an NVIDIA V100 GPU with 16 GB GPU memory.

\subsection{Results}

This section compares the performance of the proposed DeepONet solutions in approximating the dynamics of a 4th-order SM subject to time-varying voltage inputs. The objective is to assess their accuracy, generalization capability across diverse operating conditions, and computational efficiency. We benchmarked the performance against a different ML surrogate, a PINN with identical inputs, outputs, and training data, differing only in its internal network architecture. All training hyperparameters were kept identical and the number of trainable parameters was of comparable magnitude. Each experiment is repeated three times to ensure fairness, with a different random seed used for every run. The average results are reported below.

\subsubsection{Accuracy}

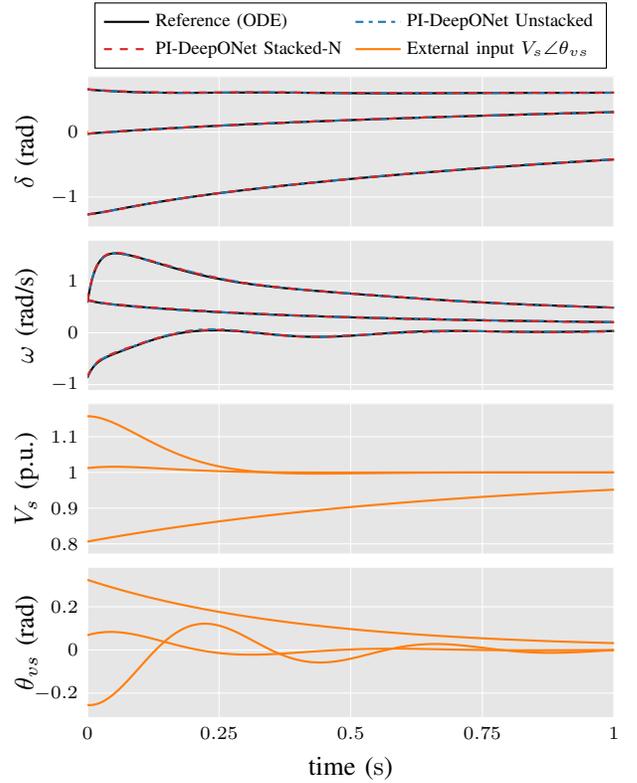
\begin{figure}[tbp]
\centering
\begin{tikzpicture}
\begin{groupplot}[
  group style={
    group name=mygroup,
    group size=1 by 4,
    xticklabels at=edge bottom,
    horizontal sep=1.1cm,
    vertical sep=5pt
  },
  ylabsh=-2.3em,
  xmin=0, xmax=1,
  xtick distance=0.25,
  width=7cm,
  height=2cm,
  legend style={
    font=\scriptsize,
    at={(0.5,1.05)},           % Position: center above plot
    anchor=south,             % Attach legend bottom to that point
    legend columns=2,         % Number of columns (adjust as needed)
    /tikz/every even column/.append style={column sep=0.1cm}
  },
  table/col sep=space,
  table/header=true
]

% 1) theta
\nextgroupplot[
  ylabel={$\delta$ (rad)},
  xlabel={time (\si{\second})}
]
  \addplot[black, thick] table[x=time, y=theta]{plots2/dat_files/DeepONet_Stacked_traj_0_targets.dat};
  \addplot[black, thick, forget plot] table[x=time, y=theta]{plots2/dat_files/DeepONet_Stacked_traj_1_targets.dat};
  \addplot[black, thick, forget plot] table[x=time, y=theta]{plots2/dat_files/DeepONet_Stacked_traj_2_targets.dat};

  \addplot[sBlue, dashdotted, thick] table[x=time, y=theta]{plots2/dat_files/DeepONet_Stacked_traj_0_predictions.dat};
  \addplot[sBlue, dashdotted, thick, forget plot] table[x=time, y=theta]{plots2/dat_files/DeepONet_Stacked_traj_1_predictions.dat};
  \addplot[sBlue, dashdotted, thick, forget plot] table[x=time, y=theta]{plots2/dat_files/DeepONet_Stacked_traj_2_predictions.dat};

  \addplot[sRed, dashed, thick] table[x=time, y=theta]{plots2/dat_files/DeepONet_Unstacked_traj_0_predictions.dat};
  \addplot[sRed, dashed, thick, forget plot] table[x=time, y=theta]{plots2/dat_files/DeepONet_Unstacked_traj_1_predictions.dat};
  \addplot[sRed, dashed, thick, forget plot] table[x=time, y=theta]{plots2/dat_files/DeepONet_Unstacked_traj_2_predictions.dat};
  \addplot[sOrange, thick] coordinates {(0,0) (0,0)};
  % === Manual legend ===
  \addlegendimage{line legend, black, thick}
  \addlegendentry{Reference (ODE)}
  \addlegendimage{line legend, sRed, dashed, thick}
  \addlegendentry{PI-DeepONet Unstacked}
  \addlegendimage{line legend, sBlue, dashdotted, thick}
  \addlegendentry{PI-DeepONet Stacked-N}
  \addlegendimage{line legend, sOrange, thick}
  \addlegendentry{External input $V_s\angle \theta_{vs}$}

% 2) omega
\nextgroupplot[
  ylabel={$\omega$ (rad/s)},
  xlabel={time (\si{\second})}
]
  % targets (3 traj)
  \addplot[black, thick] table[x=time, y=omega]{plots2/dat_files/DeepONet_Stacked_traj_0_targets.dat};
  \addplot[black, thick, forget plot] table[x=time, y=omega]{plots2/dat_files/DeepONet_Stacked_traj_1_targets.dat};
  \addplot[black, thick, forget plot] table[x=time, y=omega]{plots2/dat_files/DeepONet_Stacked_traj_2_targets.dat};

  % stacked (3 traj)
  \addplot[sBlue, dashdotted, thick] table[x=time, y=omega]{plots2/dat_files/DeepONet_Stacked_traj_0_predictions.dat};
  \addplot[sBlue, dashdotted, thick, forget plot] table[x=time, y=omega]{plots2/dat_files/DeepONet_Stacked_traj_1_predictions.dat};
  \addplot[sBlue, dashdotted, thick, forget plot] table[x=time, y=omega]{plots2/dat_files/DeepONet_Stacked_traj_2_predictions.dat};

  % unstacked (3 traj)
  \addplot[sRed, dashed, thick] table[x=time, y=omega]{plots2/dat_files/DeepONet_Unstacked_traj_0_predictions.dat};
  \addplot[sRed, dashed, thick, forget plot] table[x=time, y=omega]{plots2/dat_files/DeepONet_Unstacked_traj_1_predictions.dat};
  \addplot[sRed, dashed, thick, forget plot] table[x=time, y=omega]{plots2/dat_files/DeepONet_Unstacked_traj_2_predictions.dat};

\nextgroupplot[
  ylabel={$V_{s}$ (p.u.)},
  xlabel={time (\si{\second})}
]
  % targets (3 traj)
  \addplot[sOrange, thick] table[x=time1, y=ext11]{plots2/dat_files/branch_input_wide_format.dat};
  \addplot[sOrange, thick, forget plot] table[x=time1, y=ext12]{plots2/dat_files/branch_input_wide_format.dat};
  \addplot[sOrange, thick, forget plot] table[x=time1, y=ext13]{plots2/dat_files/branch_input_wide_format.dat};

% 3) E'_d
\nextgroupplot[
  ylabel={$\theta_{vs}$ (rad)},
  xlabel={time (\si{\second})}
]
  % targets (3 traj)
  \addplot[sOrange, thick] table[x=time1, y=ext21]{plots2/dat_files/branch_input_wide_format.dat};
  \addplot[sOrange, thick, forget plot] table[x=time1, y=ext22]{plots2/dat_files/branch_input_wide_format.dat};
  \addplot[sOrange, thick, forget plot] table[x=time1, y=ext23]{plots2/dat_files/branch_input_wide_format.dat};

% 4) Vs

\end{groupplot}
\end{tikzpicture}
\caption{Illustration of three trajectories from the test dataset showing the rotor angle $\delta$ and speed $\omega$ obtained from the reference ODE solver, the PI-DeepONet surrogates, and the single PINN model. The corresponding time-varying external voltage magnitude $V_s$ and phase $\theta_{vs}$ are shown in the lower panels.}
\label{fig:traj_examples}
\vspace{-1.5em}
\end{figure}

%The two DeepONet architectures and especially their PI-implementations accurately capture the dynamic evolution of all four state variables across the test dataset. The evaluation was conducted on 100 unseen trajectories generated from randomly sampled initial conditions and external input signals within the same input domain.
The evaluation was conducted on the 100 unseen trajectories of the test dataset, where each trajectory includes the full dynamic evolution of the four state variables. Figure.~\ref{fig:traj_examples} highlights the accurate trajectory reconstruction of the operator-learning surrogates across different initial conditions and external inputs. 

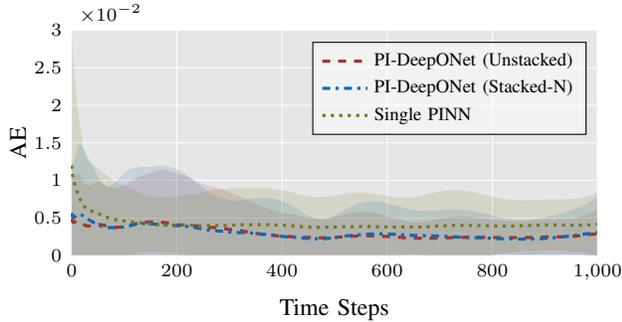
\begin{figure}[tbp]
\centering
\begin{tikzpicture}
\begin{axis}[
  width=7cm, height=3cm,
  xlabel={Time Steps},
  ylabel={AE},
  grid=both,
  grid style={line width=0.1pt, draw=gray!30},
  xmin=-1,
  ymin=0,
  % ---- Legend outside, top-right:
  legend style={
  font=\scriptsize,
    at={(0.98,0.95)},           % Position: center above plot
    anchor=north east,             % Attach legend bottom to that point
    legend columns=1,         % Number of columns (adjust as needed)
    /tikz/every even column/.append style={column sep=0.2cm},
    fill opacity = 0.5,
    text opacity = 1
  },
  ylabsh=-2em, 
  ytick distance = 0.005, 
  xmax = 1000,    % minimum and maximum of x axis
  ymax = 0.03,    % minimum and maximum of x axis
  xtick distance = 200, 
  legend cell align=left,
  font=\small,
  tick label style={/pgf/number format/fixed},
  every axis plot/.append style={line join=round},
  scaled y ticks = true,
  tick scale binop = \times, 
  axis line style={line width=0.5pt},
  major tick style={line width=0.3pt},
  minor tick style={line width=0.2pt},
  table/col sep=space,
]

% --- DeepONet Unstacked ---
% Confidence band (out of legend)
\addplot[name path=unstackhigh, draw=none, forget plot]
  table[x index=0, y index=3]{plots2/dat_files/sm4_unstacked_mae_data.dat};
\addplot[name path=unstacklow, draw=none, forget plot]
  table[x index=0, y index=2]{plots2/dat_files/sm4_unstacked_mae_data.dat};
\addplot[on layer=axis background, fill=pRed, fill opacity=0.15, draw=none, forget plot]
  fill between[of=unstackhigh and unstacklow];

% Main line
\addplot[no marks, line width=1.2pt, draw=pRed, dashed]
  table[x index=0, y index=1]{plots2/dat_files/sm4_unstacked_mae_data.dat};
% Legend sample:
\addlegendentry{PI-DeepONet (Unstacked)}
%\addlegendimage{no markers, draw=pRed, line width=1.2pt, dashed}

% --- DeepONet Stacked ---
% Confidence band (keep out of legend)
\addplot[name path=stackhigh, draw=none, forget plot]
  table[x index=0, y index=3]{plots2/dat_files/sm4_stacked_mae_data.dat};
\addplot[name path=stacklow, draw=none, forget plot]
  table[x index=0, y index=2]{plots2/dat_files/sm4_stacked_mae_data.dat};
\addplot[on layer=axis background, fill=pBlue, fill opacity=0.15, draw=none, forget plot]
  fill between[of=stackhigh and stacklow];

% Main line
\addplot[no marks, line width=1.2pt, draw=pBlue, dashdotted]
  table[x index=0, y index=1]{plots2/dat_files/sm4_stacked_mae_data.dat};
% Legend sample to match exactly:
\addlegendentry{PI-DeepONet (Stacked-N)}
%\addlegendimage{no markers, draw=pBlue, line width=1.2pt, solid}

% --- Single PINN ---
% Confidence band (out of legend)
\addplot[name path=pinnhigh, draw=none, forget plot]
  table[x index=0, y index=3]{plots2/dat_files/sm4_pinn_mae_data.dat};
\addplot[name path=pinnlow, draw=none, forget plot]
  table[x index=0, y index=2]{plots2/dat_files/sm4_pinn_mae_data.dat};
\addplot[on layer=axis background, fill=pGreen, fill opacity=0.15, draw=none, forget plot]
  fill between[of=pinnhigh and pinnlow];

% Main line
\addplot[no marks, line width=1.2pt, draw=pGreen, dotted]
  table[x index=0, y index=1]{plots2/dat_files/sm4_pinn_mae_data.dat};
% Legend sample:
\addlegendentry{Single PINN}
%\addlegendimage{no markers, draw=pGreen, line width=1.2pt, dotted}

\end{axis}
\end{tikzpicture}\vspace{-0.4em}
\caption{Comparison of the absolute error (AE) over time among the three surrogate models, illustrating their relative accuracy and temporal stability.}
%Comparison of the absolute error (AE) over time among the three surrogate models. The solid lines indicate the mean AE across 100 test trajectories, while the shaded regions represent the standard deviation of the AE, illustrating their relative accuracy and temporal stability.
\label{fig:accuracy_results}
\vspace{-1.2em}
\end{figure}

Based on  the accuracy metrics summarized in Table~\ref{accu_table}, among all variations, the Unstacked PI-DeepONet achieved the best overall performance, followed by the Stacked-N PI-DeepONet, while all DeepONet-based models significantly outperformed the baseline PINN. 
The temporal evolution of the AE is illustrated in Figure~\ref{fig:accuracy_results}
Both PI-DeepONet architectures exhibit slightly higher errors during the initial transient period but remain stable and low throughout the remainder of the time horizon. In contrast, the PINN excibits consistently higher errors and demonstrates weaker performance, particularly in the transient region.
Overall, these results confirm the superior capability of PI-DeepONets to learn functional mappings between input and output trajectories with high accuracy and robustness.
It is worth noting that, although not presented in this work, additional experiments indicate that increasing the number of training samples consistently enhances the accuracy of all implementations.

\begin{table}[tbp]
\centering
\caption{Average accuracy metrics of ML surrogates.}
\label{accu_table}
% \scriptsize
\centering
\setlength{\tabcolsep}{3pt} % Adjust column separation
\renewcommand{\arraystretch}{0.9} % Adjust row separation
\begin{tabular}{p{2cm}ccc}
\toprule
\textbf{ ML Surrogate} & \textbf{MSE} & \textbf{MAE} & \textbf{Max AE}  \\
\midrule
 PINN                   & $38.67 \times 10^{-6} $ & $3.87 \times 10^{-3}$ & $172.93\times 10^{-3}$\\\midrule
  Unstacked\newline DeepONet & $38.25 \times 10^{-6} $ & $3.66 \times 10^{-3}$ & $117.23 \times 10^{-3}$\\
  \midrule
 Unstacked \newline PI-DeepONet & $26.87 \times 10^{-6} $ & $3.26 \times 10^{-3}$ & $73.03 \times 10^{-3}$\\
 \midrule
 Stacked-N \newline DeepONet & $38.6 \times 10^{-6} $ & $3.68 \times 10^{-3}$ & $157.44 \times 10^{-3}$\\
  \midrule
 Stacked-N \newline PI-DeepONet & $36.14 \times 10^{-6}$ & $ 3.36 \times 10^{-3}$  & $104.69 \times 10^{-3}$\\
\bottomrule

\end{tabular}
\vspace{-0.3em}
\end{table}

\subsubsection{Time}

In addition to accuracy, computational efficiency is a critical factor for enabling real-time applications such as online stability assessment and digital twins. Once trained within a specified input domain, with an upfront training cost of up to 15 minutes, the DeepONets can infer complete system trajectories without requiring step-by-step numerical integration. Table~\ref{time_table} summarizes the runtime comparison: for a single trajectory, the DeepONet achieves more than a $30\times$ speedup compared to a high-precision Runge--Kutta solver, while the advantage becomes even more pronounced when evaluating multiple trajectories in parallel, reaching a speedup exceeding $6000\times$ for 1000 trajectories. It should be noted that the ODE solver runs sequentially on a single CPU core, whereas DeepONet inference leverages parallel GPU execution through a single forward evaluation. In contrast, PINNs remain slower due to their dense architecture and large number of neuron connections, which increase computational and memory costs. Overall, these results highlight the scalability and suitability of the proposed approach for real-time applications.
% \begin{table}[t]
% \centering
% \caption{Inference time in ms (and Speedup)}
% \label{time_table}
% \scriptsize
% \centering
% \setlength{\tabcolsep}{3pt} % Adjust column separation
% \renewcommand{\arraystretch}{0.9} % Adjust row separation
% \begin{tabular}{l S[table-format=3.3]S[table-format=3.3]S[table-format=3.3]}
% \toprule
%   & \textbf{Single Trajectory} & \textbf{10 Trajectories} & \textbf{1000 Trajectories} \\
% \midrule
% \textbf{ODE solver} & 15.517  & 155.169  & 15516.834 \\\midrule
% \textbf{PINN} & 0.189
%  &  0.665 & 6.861 \\\midrule
% \textbf{Unstacked DeepONet} & 0.358
%  &  0.425 & 2.311 \\\midrule
% \textbf{Stacked-N DeepONet} & 0.485 &  0.539 & 2.415 \\
% \bottomrule
% \end{tabular}
% \end{table}

\begin{table}[t]
\centering
\caption{Inference time in ms (and Speedup)}
\label{time_table}
\scriptsize
\centering
\setlength{\tabcolsep}{3pt} % Adjust column separation
\renewcommand{\arraystretch}{0.9} % Adjust row separation
\begin{tabular}{l c c c}
\toprule
 & \textbf{Single Trajectory} & \textbf{10 Trajectories} & \textbf{1000 Trajectories} \\
\midrule
\textbf{ODE solver} & 15.517 & 155.169 & 15516.834 \\
\midrule
\textbf{PINN} & 0.189 ($ 80\times$) & 0.665 ($ 230\times$) & 6.861 ($ 2260\times$) \\
\midrule
\textbf{Unstacked DeepONet} & 0.358 ($ 40\times$) & 0.425 ($ 360\times$) & 2.311 ($ 6720\times$) \\
\midrule
\textbf{Stacked-N DeepONet} & 0.485 ($ 30\times$) & 0.539 ($ 290\times$) & 2.415 ($ 6430\times$) \\
\bottomrule

\end{tabular}
\vspace{-1.8em}
\end{table}

\section{Conclusion and Future Work}
This paper introduced physics-informed neural operators for modeling of power system dynamic components. By combining the capabilities of the DeepONet's architecture with physics-informed training, the proposed approach enables accurate and efficient trajectory prediction for systems subject to time-varying inputs. Once trained, the operator can evaluate entire trajectories in a single forward pass, bypassing the need for iterative integration and achieving substantial speedups compared to conventional solvers. The results demonstrate that embedding physics into the learning process enhances generalization to previously unseen operating conditions.

Future research will focus on improving the scalability and adaptability of neural operator models. This includes investigating sensor placement strategies and search for optimal operator hyperparameters, developing uncertainty quantification and adaptive loss weighting techniques, and extending the framework to encompass inverter-based resources and multi-machine systems. Integrating neural operators into real-time simulation and control frameworks, together with advancing online learning from measurement data, constitutes a significant step toward developing fully data- and physics-informed digital twins of the power grid.
% The 'triggered' command can be changed if desired:


%\bibliographystyle{IEEEtran}

\bibliographystyle{IEEEtran}
\bibliography{references_paper_arxiv}

\end{document}